\documentclass[%
reprint,
superscriptaddress,
frontmatterverbose, 
 amsmath,amssymb,
aps,
]{revtex4-2}
\usepackage{xcolor}
\usepackage{graphicx}
\usepackage{dcolumn}
\usepackage{bm}
\usepackage{blindtext}
\usepackage{hyperref}
\usepackage[mathlines]{lineno}

\usepackage[
scale=0.85, marginratio={1:1, 2:3}, ignoreall,
margin=0.65in,
]{geometry}

\begin{document}

\preprint{APS/123-QED}

\title{Direct Hydrogen Quantification in High-pressure Metal Hydrides}

  \author{Thomas Meier}
  \email{thomas.meier@hpstar.ac.cn}
  \affiliation{Center for High Pressure Science and Technology Advance Research, Beijing, China}

  \author{Dominique Laniel}
  \affiliation{Center for Science at Extreme Conditions, University of Edinburgh, Edinburgh, UK}  
   
  \author{Florian Trybel}
  \affiliation{Department of Physics, Chemistry and Biology (IFM), Link{\"o}ping University, Link{\"o}ping, Sweden} 
  
\date{\today}

\begin{abstract}
High-pressure metal-hydride (MH) research evolved into a thriving field within condensed matter physics following the realisation of metallic compounds showing phonon mediated near room-temperature superconductivity. However, severe limitations in determining the chemical formula of the reaction products, especially with regards to their hydrogen content, impedes a deep understanding of the synthesized phases  and can lead to significantly erroneous conclusions. Here, we present a way to directly access the hydrogen content of MH solids synthesised at high pressures in (laser-heated) diamond anvil cells using nuclear magnetic resonance (NMR) spectroscopy. We show that this method can be used to investigate MH compounds with a wide range of hydrogen content, from MH$_x$ with x=0.15 (CuH$_{0.15}$) to $x\lesssim 6.4$ (H$_{6\pm 0.4}$S$_5$).
\end{abstract}
\maketitle
\section{Introduction}

\begin{figure*}[t]
\includegraphics[width=0.8\textwidth]{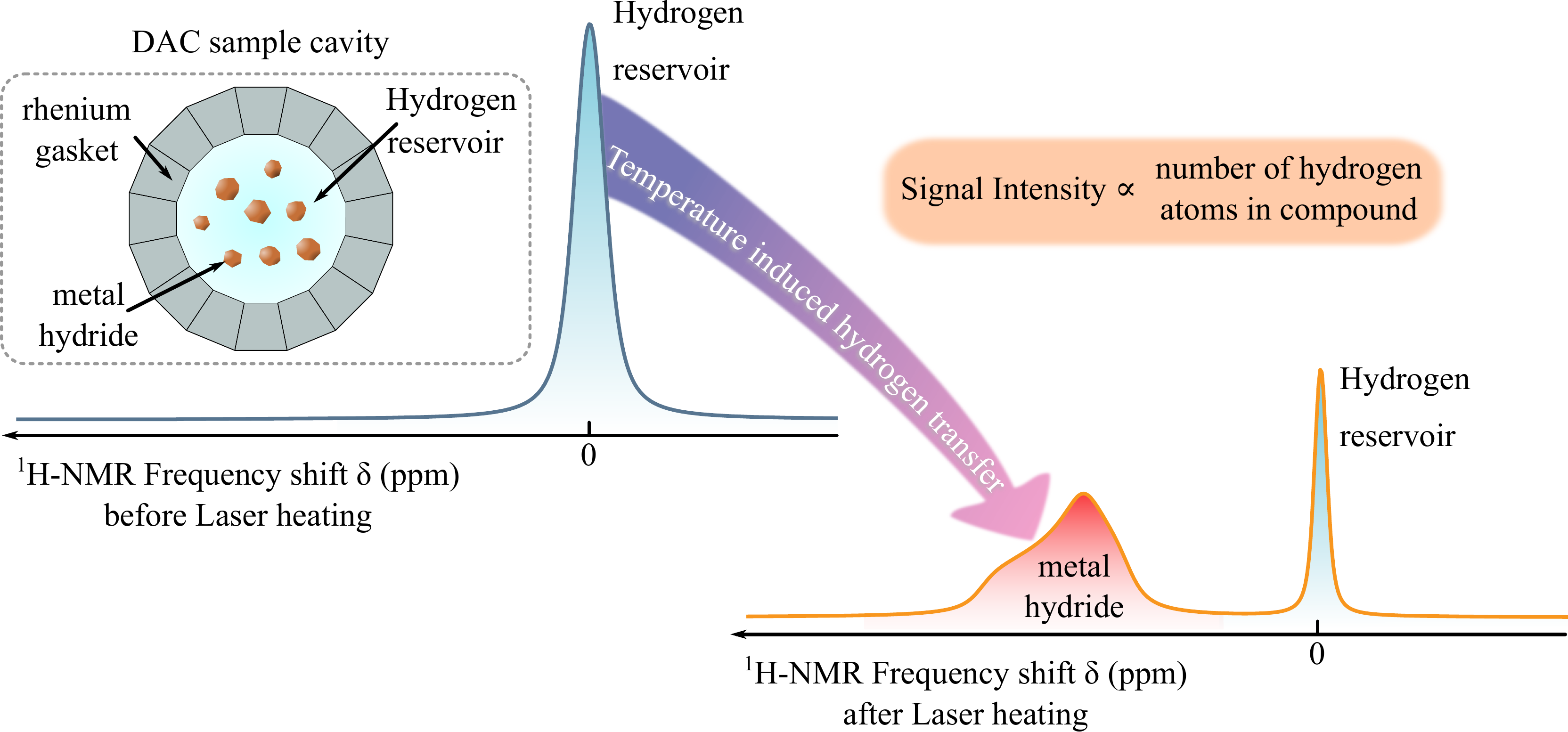}
\caption{Schematic representation of the quantification method. Prior to laser heating, the sample cavities are filled with powder or larger crystals of the parent metals (e.g. Cu, Fe, Y, S etc.) as well as a suitable hydrogen reservoir like molecular hydrogen, paraffin or ammonia borane. The resulting $^1$H-NMR spectra solely contain signals stemming for the hydrogen rich precursors. Surpassing thermal barriers via laser heating, hydrogen diffusion into the parent metals will facilitate the formation of metal hydrides whose $^1$H-NMR signal intensity (i.e. signal-to-noise ratios) will be proportional to the amount of hydrogen atoms in the hydride.}
\label{fig:principle}
\end{figure*} 

The theoretical prediction of high-temperature superconductivity in pressure-stabilised hydrogen-rich metal hydrides (MH) \cite{Struzhkin2020a, Ashcroft2004, Zurek2009}, and their subsequent experimental confirmation \cite{Drozdov2015, Drozdov2019, Kong2019, Somayazulu2018}, is considered the dawn of one the most prolific research fields within the high-pressure community in the past decades \cite{Flores-Livas2019}. 
\\
The fruitful synergy between \textit{ab-initio} structure search methods \cite{Oganov2008, Pickard2011, Wang2012}, 
 and electronic calculations \cite{Pickard2020, Errea2016} with experimental characterization from synchrotron-based X-ray diffraction (XRD) and transport methods, forms a well-established framework for the search of novel binary or ternary metal hydride compounds with superconducting transition temperatures ($T_{c}$) rapidly approaching 300 K \cite{Snider2020}.
\\
The hydrogen subsystems play a crucial role in the realisation of high superconducting transition temperatures in pressure-formed metallic hydride solids \cite{Errea2020,Belli2021}, therefore a precise knowledge of the content, atomic position, electronic and vibrational properties of a metal hydride's H-sublattice is of the utmost importance for a detailed understanding of the underlying physical phenomena governing this new material class.  
\\
Directly accessing the hydrogen subsystem with standard high-pressure methods has been, unfortunately, unfeasible. Indeed, hydrogen atoms are essentially invisible to XRD methods \cite{Ji2020a, Gregoryanz2020} due to having a single electron. While neutron diffraction does not suffer from this, it cannot routinely be used at the pressures required for the stabilization of most high $T_{c}$  materials \cite{Boehler2013}. Moreover, methods such as Raman and infrared spectroscopy greatly struggle with metallic samples. 
\\
Thus, the characterization of metallic hydrides is often limited to a structural investigation of the much heavier parent metal lattices and their actual hydrogen content can only be approximately inferred from the expansion of the hydrides' unit cell upon hydrogen uptake. The hydrogen positions can then be obtained by comparison with structures found stable in \textit{ab-inito} calculations \cite{Binns2019, Pena-Alvarez2019, Pepin2017}. However, there are various limitations of this approach: (i) The MHs of interest can in most cases not be synthesised phase pure. Therefore, an unambiguous identification of the space group is usually difficult, especially from powder diffraction patterns. (ii) Relating the volume expansion of the unit cell to the hydrogen content relies on knowledge about volume per hydrogen as well as the other atoms at a certain pressure and (synthesis) temperature in the respective chemical environment, which can only be approximated to a certain accuracy from analogue experiments or the respective equations of state.  
(iii) The calculated unit cell volumes as well as structural and dynamical stability can be influenced by a variety of parameters and especially for hydrides thermal as well as nuclear quantum effects can be crucial for finding the correct structure \cite{Errea2016}, but are widely not taken into account.
\\
In summary, there is a considerable potential for errors and therefore large deviations of the estimate from the actual hydrogen content of the synthesized hydride(s). Quantitative nuclear magnetic resonance (NMR) spectroscopy methods, on the other hand, are a well established analytical tool in modern laboratories \cite{Simmler2014, Giraudeau2017, Evilia2001} and especially suited for hydrogen related experiments. Moreover, recent developments in \textit{in-situ} high-pressure NMR in diamond anvil cells (DACs) at pressures up to 200 GPa \cite{Meier2017, Meier2018a, Meier2019a} present an outstanding opportunity for the direct observation of hydrogen atoms the relevant high-pressure conditions. 
\\
Here, we introduce a method suitable for the direct quantification of the hydrogen content in metal hydrides for a wide range of compositions and hydrogenations spanning an order of magnitude, based on the inherent relationship between signal intensity and number density of resonant nuclei \cite{Meier2017b, Meier2018b} in \textit{in-situ} diamond anvil cell NMR experiments at extreme pressures. 

\section{Experimental Details}

We investigated seven metal hydrides with different hydrogen contents:  the copper hydrides CuH$_{0.15}$, Cu$_2$H and CuH, iron monohydride, the yttrium hydrides YH$_2$ and YH$_3$ as well as the recently discovered tetragonal phase $tI140$ of sulfur polyhydride (H$_{6\pm \delta}$S$_5$) ($\delta\approx 0.4$) by Laniel et al.\cite{Laniel2020a}. 

DAC preparation for iron and copper hydrides are described elsewhere \cite{Meier2020a,Meier2019a}. Pressure cells for sulfur (S-cell) and yttrium (Y-cell) hydrides were prepared in a similar fashion. Small pieces of elemental sulfur as well as yttrium metal were loaded into a DAC set-up for NMR experiments and loaded with paraffin oil as pressure transmitting medium as well as hydrogen reservoir.  

The cells were precompressed to 45 GPa (Y-cell) and 65 GPa (S-cell) and their respective $^1$H-NMR spectra were recorded using a magnetic field of 7.04 T, corresponding to 300 MHz $^1$H-NMR frequencies. Pulse nutation experiments were used to determine optimal $\pi/2$-pulse excitation conditions with long scan repetition times to allow for sufficient relaxation of the pressurised paraffin spin system (spin-lattice relaxation time $T_1~\approx~750$ ms). 

Afterwards, both cells were laser-heated with a double-sided laser heating system \cite{Fedotenko2019} to about 1500 to 2000 K. Corresponding $^1$H-NMR spectra after laser heating featured additional signals which were associated with the formation of YH$_3$ and YH$_2$ in the Y-cell (\textit{c.f.} \cite{Meier2021a}) and H$_{6\pm 0.4}$S$_5$ (Fig. \ref{fig:spectra}b). 

Data analysis was performed using the ONMR plugin in Origin 2019 Pro.

\section{Quantification method}
\subsection{Background}

In order to quantify the hydrogen content in the metal hydride compounds, we take advantage of the signal strengths in NMR experiments being directly proportional to the amount of resonating spins contributing to the signal. The scan-normalised signal-to-noise ratio ($\zeta$), after single pulse excitation for infinitely short radio-frequency (RF) pulses,  is given by \cite{Meier2017a}: 

\begin{equation}
    \zeta=\frac{{\rm SNR}}{\sqrt{N_{\rm scans}}}= \frac{\eta N_{\rm coil} A_{\rm coil} \omega_0 \mu_0 M_0}{\sqrt{4R_{ \rm coil} k_{\rm B} T \Delta f}}
    \label{eq:SNR},
\end{equation}
where $\zeta$ is the scan-normalized signal-to-noise ratio (SNR), $N_{\rm scans}$ the number of accumulations, $\eta$ the RF-coils' filling factor (i.e. the ratio between sample and resonator volume), $N_{\rm coil}$ and $A_{\rm coil}$ are the resonators' number of windings and cross-section, respectively. $\omega_0$ is  the resonance frequency in angular units, which directly relates to the magnetic field strength via $\omega_0=\gamma_{\rm n} B_0$ with $\gamma_{\rm n}$ being the isotope-specific gyromagnetic ratio. The thermal Nyquist noise term - the  denominator of equation \eqref{eq:SNR} -  is given by the resonators' temperature $T$, the limiting bandwidth $\Delta f$ (usually the inverse dwell time, or digitisation frequency) and the AC resistance of the coil $R_{\rm coil}$ (corrected for limited penetration of the RF current in the conductive resonator structures) and the Boltzmann constant $k_{\rm B}$.

The nuclear magnetisation $M_0$ in equation \eqref{eq:SNR} is given by the Curie-law for nuclear spins \cite{Abragam1961}:

\begin{equation}
   M_0=n\cdot\frac{\gamma_{\rm n}^2 \hbar^2 I(I+1) B_0}{3k_{\rm B} T}
    \label{M0}
\end{equation}
with the nuclear spin quantum number $I$, the external magnetic field $B_0$ and the prefactor $n=N/V$ representing the concentration of the nuclear spins contributing to the observed NMR signal.  


For a full quantification of the hydrogen content in the reaction products after laser heating three assumptions have to be made, both well justified given the experimental conditions: 

i) Equation \ref{eq:SNR} is valid within the approximation of infinitely short RF-pulses, i.e. partial spin relaxation within the RF-excitation is negligible. This assumption is exceptionally well met in \textit{in-situ} high pressure NMR experiments where resonator structures are particularly small, leading to strong exciting RF magnetic fields and thus allow for excitation pulses in the order of 1 $\mu s$ or lower.

ii) All geometric coil parameters given in equation \ref{eq:SNR} do not change significantly upon laser heating and signal intensities are directly proportional to (1) the nuclear spin density $n$, (2) the filling factor of the compound $\eta$ and (3) the DAC's cavity volume through the SNR:
\begin{equation}
     \zeta \propto n\cdot\eta
\end{equation}

iii) the system is closed, i.e. the total hydrogen content in the DAC's sample chamber remains constant, and the full experimental cavity is always probed, such that:

\begin{align}
N^0_{\rm Res}&=N^{\rm LH}_{\rm Res}+\sum_{i} N^{\rm LH}_{\rm MH,i},\\
\eta^0_{\rm Res}&=\eta^{\rm LH}_{\rm Res}+ \sum_{i} \eta^{\rm LH}_{\rm MH,i}, 
\label{eq:heating}
\end{align}
where $N^0_{\rm Res}$ and $\eta^0_{\rm Res}$ denote the number of nuclei contributing to the respective signal and the filling factor of the signal-generating compound within the DAC's sample cavity before laser heating. $N^{\rm LH}_{\rm Res/MH,i}$ and $\eta^{\rm LH}_{\rm Res/MH,i}$  denote the number of nuclei and filling factor of the hydrogen reservoir (Res) and $i$th metal hydride (MH,i) after laser heating, respectively. 

Solving these equations for the hydrogen density in the reaction product, $n^{\rm LH}_{\rm MH,i}$, yields:

\begin{equation}
    n^{\rm LH,i}_{\rm MH}=n^0_{\rm Res} \cdot \frac{ \zeta^{\rm LH}_{\rm MH,i}}{ \zeta^0_{\rm Res}- \zeta^{\rm LH}_{\rm Res}-\sum_{j \neq i}\zeta^{\rm LH}_{\rm MH,j}}.
    \label{n1}
\end{equation}

\begin{figure}[htb]
\begin{center}
\includegraphics[width=0.45\textwidth]{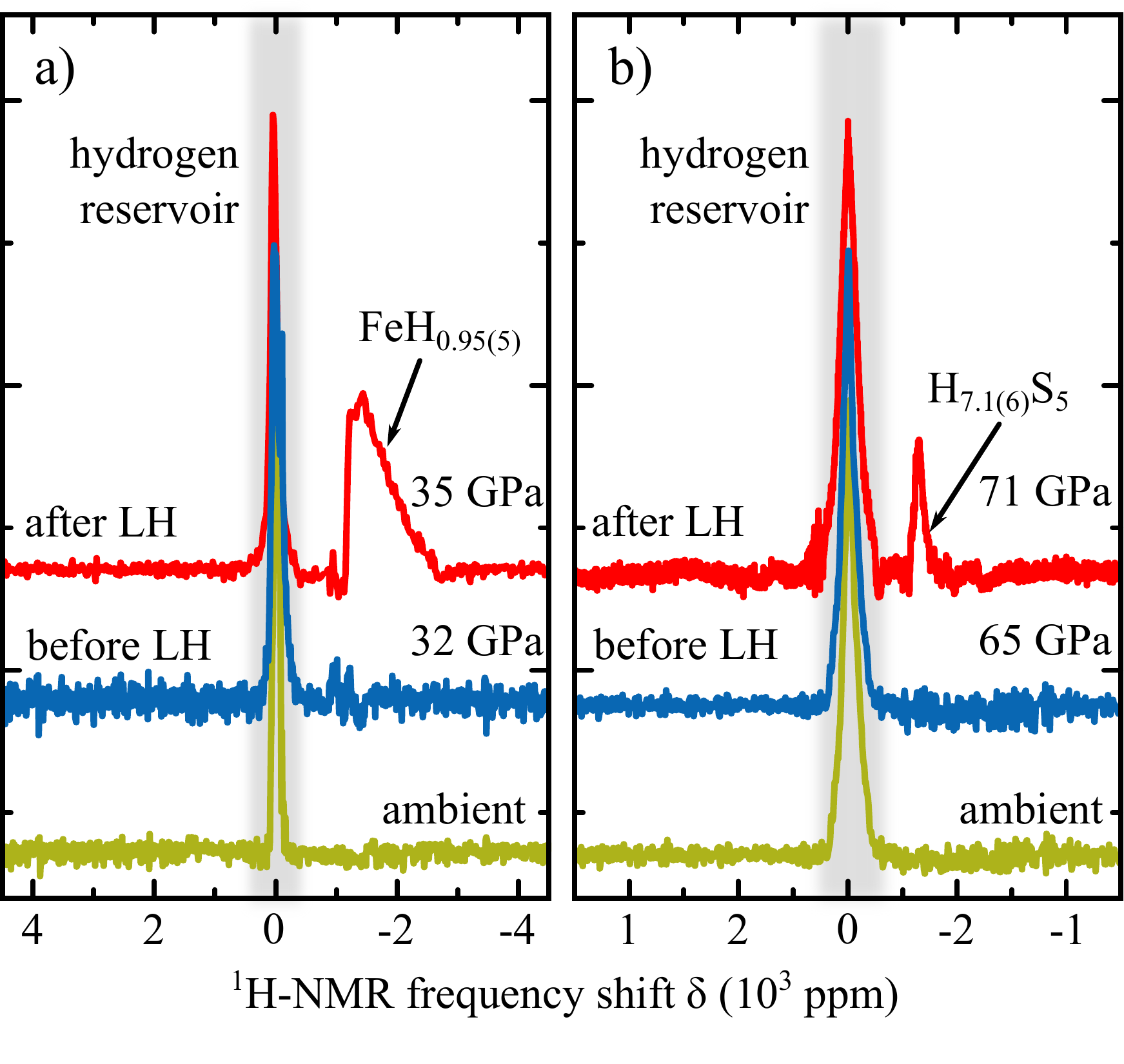}%
\caption{Representative $^1$H-NMR spectra used for the quantification of the metal hydrides' hydrogen content. \textbf{a) Iron hydride:} A strong additional resonance was found in the NMR spectra which was identified to be \textit{fcc} iron hydride, see \cite{Meier2019a}. comparison of signal-to-noise ratios resulted in a hydrogen content of $x=0.95(5)$ H-atoms per formula unit, close to the inferred hydrogen concentration from XRD measurements. Spectra are taken from earlier studies \cite{Meier2019a}. \textbf{b) Sulfur hydride:} Laser heating in the DAC was conducted at 65 GPa (blue spectrum), resulting in the occurrence of a weak additional signal at about -1750 ppm at 71 GPa containing approximately 177 H-atoms per unit cell, corresponding to a hydrogenation of $x=7.1(9)$ H-atoms per formula unit in good correspondence with the recently reported tetragonal $tI104$ phase of H$_{6\pm0.4}$S$_5$  by Laniel et al.\cite{Laniel2020a}.}
\label{fig:spectra}
\end{center}
\end{figure} 

In order to ensure reproducible results for these quantitative NMR experiments, proper excitation of all contributing spin systems - using a well-defined RF excitation pulse - as well as sufficiently long pulse repetition times - to allow for full relaxation of all spin systems - needs to be ensured. 

\subsection{Examples}

In the following section, we present a detailed step-by-step procedure for the hydrogen quantification experiments of two representative hydrides: iron monohydride and sulfur polyhydride. The same steps were applied to all other investigated compounds, for which the necessary parameters for the calculation of $n^{\rm LH,i}_{\rm MH}$ are given in the supplementary material.

\subsubsection{Iron monohydride}

Figure \ref{fig:spectra}a) shows $^1$H-NMR spectra from a DAC using iron powder and paraffin oil as a hydrogen reservoir before (yellow spectrum at ambient conditions, blue spectrum at 32 GPa) and after laser heating (red spectrum). Both the ambient and the spectra after laser heating were recorded at a spectrometer frequency of $f_0=45.0831$ MHz ($B_0=1059$ mT) with digitisation frequencies of 1 and 3.33 MHz and 1224 and 1988 scans for the spectra before and after laser heating, respectively. 

The corresponding SNR, corrected for a differing number of scans and digitisation frequencies (see eq.\ref{eq:SNR}), of the paraffin oil signal before and after heating as well as of the hydride signal were found to be $\zeta^{\rm 0}_{\rm Res}=23674$, $\zeta^{\rm LH}_{\rm Res}=17158$ and $\zeta^{\rm LH}_{\rm FeH_{\rm x}}=7702$, respectively. Assuming a hydrogen density of the precursor of $n^{\rm 0}_{\rm Res}=0.0729(73)~$H/\AA$^3$\cite{Freund1982, Dorset1995, ThermoFisherScientific2012}, the hydrides' hydrogen density can be directly calculated to be $n^{\rm LH}_{\rm FeH_{\rm x}}=0.086(18)~$H/\AA$^3$. Using the equation of state from Narygina et al.\cite{Narygina2011}, the hydrogen concentration directly after the compound's synthesis at 35 GPa amounts to 3.9(3) H/unit cell and thus gives rise to about 0.95(5) H/formula unit, reasonably close to the inferred hydrogen content from X-ray diffraction experiments for the iron monohydride ($x=1$). 

\begin{figure}[htb]
\begin{center}
\includegraphics[width=0.45\textwidth]{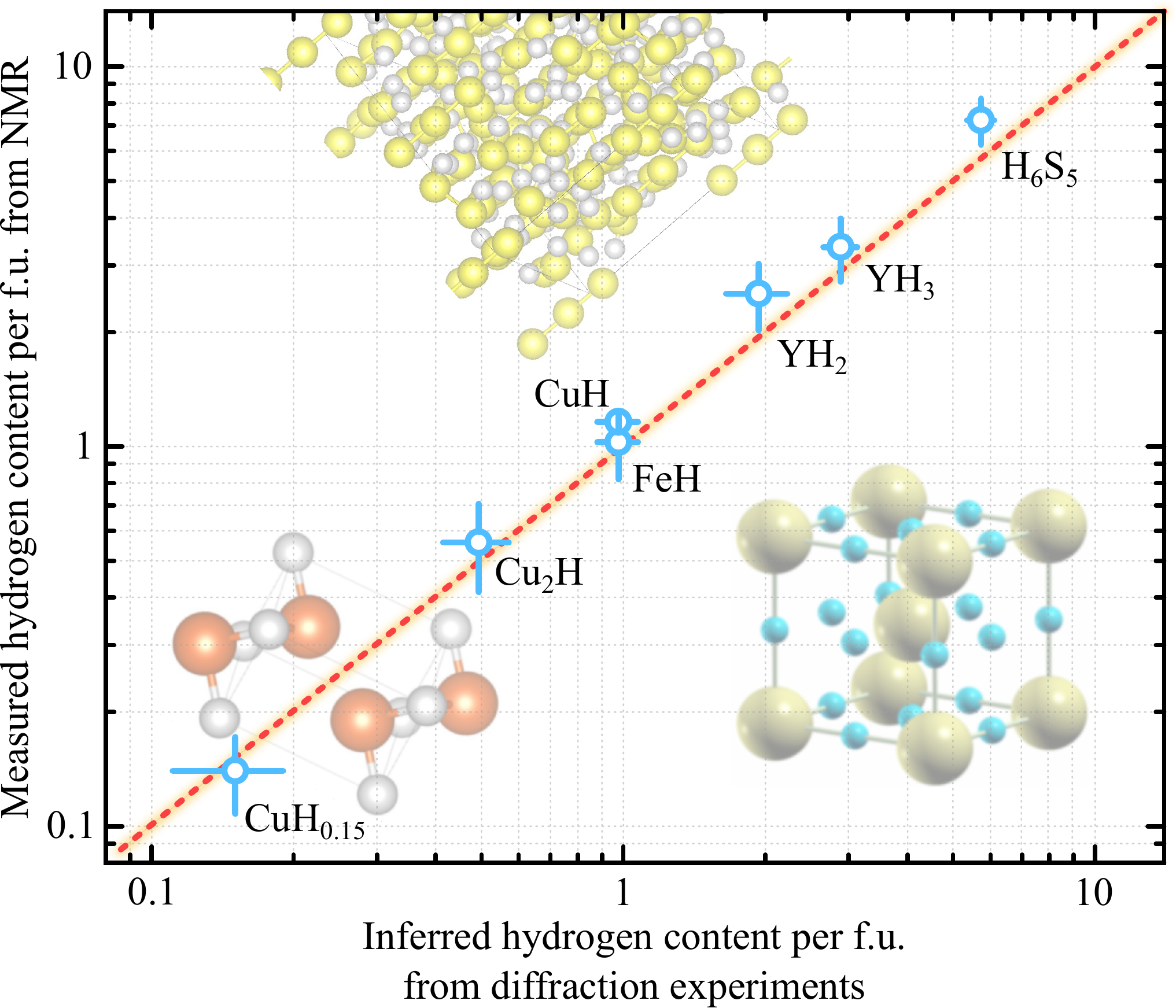}%
\caption{Comparison between inferred hydrogen content from diffraction methods and direct measurements using NMR spectroscopy in DACs. Several hydride systems have been investigated using the described method, ranging from hydrogen contents of $x=0.15$ to $x=7.1$. The red dotted line is a guide to the eye of a 1:1 correlation between both quantification methods. The structures in the background depict the low pressure trigonal Cu$_2$H, the face-centered cubic FeH as well as the tetragonal $tI$140 phase of H$_{6\pm 0.4}$S$_5$. }
\label{fig:quant}
\end{center}
\end{figure}

\subsubsection{Sulfur polyhydride}

Figure \ref{fig:spectra}b) shows recorded $^1$H-NMR spectra using sulfur and paraffin oil (yellow), pressurised to 65 GPa (blue) and laser-heated above 1500 K at $\sim 71$ GPa (red). The spectrum before laser heating at ambient conditions was recorded at a magnetic field of 7.03 T using a digitisation frequency of 2 MHz, yielding $\zeta^{\rm 0}_{\rm Res}=54685$. After laser-assisted sample synthesis, a new spectrum was recorded at a 3.33 MHz digitisation frequency, yielding re-normalised values of $\zeta^{\rm LH}_{\rm Res}=47619$ and $\zeta^{\rm LH}_{\rm H_{\rm x}S_5}=8532$. The corresponding hydrogen density of the reaction product can be estimated to be $n^{\rm LH}_{\rm H_{\rm x}S_5}=0.087(12)~$H/\AA$^3$, yielding a stochiometric hydrogen concentration of $x=7.1(9)$ H per formula unit. 

\section{Results and discussion}

The combined data of the NMR-inferred hydrogen content is displayed in figure \ref{fig:quant} in comparison with estimates from XRD and computations \cite{Burtovyy2004, Binns2019, Narygina2011, Liu2017a, Kong2021, Laniel2020a}. It can be clearly seen that directly measuring the hydrogen concentration per formula unit using NMR coincides remarkably well with the inferred hydrogenation from X-ray diffraction experiments for a wide range of stoichiometries.

Uncertainties in the obtained hydrogen content (Fig. \ref{fig:quant}) are primarily determined by the absolute signal-to-noise ratio. Using a conservative estimation, an error range of about $5 \%$ per integrated signal intensity can be considered adequate. Thus, sufficiently averaged noise patterns and strong signal intensities are favourable to ensure best accuracy in $n^{\rm LH}_{\rm Res}$.

Another possible source of inaccuracy is the hydrogen density of the precursor material. If a precursor with a know equation of state is used, it is sufficient to measure directly before the laser heating assisted sample synthesis in a precompressed DAC. In the case of an unknown or poorly constrained equation of state, the initial measurements need to be conducted at almost ambient conditions for an accurate estimation of $n^{\rm 0}_{\rm Res}$, or preliminary EoS experiments, or otherwise \emph{ab-initio} calculations, need to be performed beforehand.

Considering these points, the here presented NMR-based quantification method is in very good agreement with estimates based on X-ray diffraction and computations for a wide range of hydrogen content and further improvement in the SNR of  the NMR data accuracy will further improve the accuracy and enable the analysis of previously unknown and re-evaluation of so far non-unambiguously determined hydrogen contents. 

It is important to note that in the most general case, the here presented approach solely relies on the knowledge of the precursor material's H-atom density $n^{\rm 0}_{\rm Res}$. All other parameters are sufficiently determined by NMR. Knowledge of unit cell volumes or stoichiometries of the reaction products are not mandatory and NMR-determined hydrogen concentrations are directly given in units of H/\AA$^3$.

In special cases, this method would also allow the direct estimation of total stoichiometries, if all components of the reaction products are observable. In this vein, hydrides of Lanthanum, via $^{139}$La-NMR, and Yttrium, via $^{89}$Y-NMR, in conjunction with $^1$H-NMR, could be directly chemically characterised upon their synthesis by conducting multi-nuclear NMR experiments. In these special instances, concentrations of metal atoms per \AA$^3$ can be obtained using the same procedure as described above. The spin density $n^{\rm 0}_{\rm PM}$ and signal-to-noise ratios $\zeta^{\rm 0,LH}_{\rm PM}$ would then be given by the parent metal compounds (PM) and their respective signal strengths prior to and after sample synthesis. Furthermore, detailed knowledge of the parent metal equation-of-state and unit cell volumes are mandatory. Moreover, since different spectral ranges will be employed (e.g. $\approx400$ MHz for $^1$H and $\approx56.5$ MHz for $^{139}$La respectively), association of signals and their corresponding reaction products could be realised using two-dimensional NMR spectroscopy (e.g. hetero-nuclear correlation spectroscopy ($^1$H-$^{139}$La-COSY)).      

This independent method opens the opportunity to identify metal hydride systems on the fly with \textit{in-situ} nuclear magnetic resonance experiments under high pressures, enabling the quantification of hydrogen uptake with or without laser heating and will allow for novel characterisation experiments of this new class of high-temperature superconductors. 

Moreover, this and similar methods might be usable for other high-pressure problems like the water transport in the inner regions of Earth\cite{Hu2021a} using \textit{in-situ} high-pressure NMR experiments \cite{Trybel2020, Trybel2021, Meier2022a}

\section*{Acknowledgement and Funding}
TM acknowledges support from the Center for High Pressure Science and Technology Advance Research, Beijing, P.R. China. D.L. thanks the Alexander von Humboldt Foundation, the Deutsche Forschungsgemeinschaft (DFG, project LA-4916/1-1) and the UKRI Future Leaders Fellowship (MR/V025724/1) for financial support. 
FT acknowledges support from the Swedish Research Council (VR) grant no. 2019-05600.

%

\clearpage

\onecolumngrid 

\section*{Supplementary}
Following are the data tables used to compute the hydrogen concentrations $n^1_1$. Noise describes the root mean square noise level in spectral domain far off the resonances. Signal denotes the integral value of the signals. $N_{scans}$ are the number of scans. $f_0$ is the respective spectrometer frequency and $\Delta f$ the digitisation frequency, i.e. the limiting band-width. The last row denoted the scan normalised SNR ratio after correction for differing $f_0$ and $\Delta f_0$.

\subsection*{$FeH$}

\centering
\begingroup
\setlength{\tabcolsep}{12pt} 
\renewcommand{\arraystretch}{2.5} 
\begin{tabular}{l|l|l|l}
 & Ambient  & \multicolumn{2}{l}{{After LH}} \\
 \hline
 &  Precursor &  Precursor          &     Hydride     \\
 \hline
noise & $1.78\cdot10^4$ &    $3.01\cdot10^4$       &   $3.01\cdot10^4$       \\
signal & $1.47\cdot10^{10}$ &      $4.19\cdot10^{11}$     &     $1.88\cdot10^{10}$     \\
$N_{scans}$ & 1224 &    1988       &    1988      \\
$f_0~(MHz) $ & 45.0831 &    45.0831       &     45.0831     \\
$\Delta f ~(MHz)$ & 1 &     3.33      &     3.33     \\
$\left( \frac{SNR}{\sqrt{N_{scans}}} \right)_{rn}$ & 23674 &    17158       & 7702      
\end{tabular}
\endgroup

\subsection*{$Cu_2H$}
\centering
\begingroup
\setlength{\tabcolsep}{12pt} 
\renewcommand{\arraystretch}{2.5} 
\begin{tabular}{l|l|l|l}
 & Ambient  & \multicolumn{2}{l}{{After LH}} \\
 \hline
 &  Precursor &  Precursor          &     Hydride     \\
 \hline
noise & $1.21\cdot10^4$ &    $1.95\cdot10^5$       &   $1.95\cdot10^5$       \\
signal & $7.338\cdot10^{9}$ &      $3.11\cdot10^{11}$     &     $4.2\cdot10^{10}$     \\
$N_{scans}$ & 2048 &    25600       &    25600      \\
$f_0~(MHz) $ & 45.02 &    45.072       &     45.072     \\
$\Delta f ~(MHz)$ & 1 &     0.33      &     0.33     \\
$\left( \frac{SNR}{\sqrt{N_{scans}}} \right)_{rn}$ & 13400 &    9967       & 2333        
\end{tabular}
\endgroup

\subsection*{$CuH$}
\centering
\begingroup
\setlength{\tabcolsep}{12pt} 
\renewcommand{\arraystretch}{2.5} 
\begin{tabular}{l|l|l|l}
 & Ambient   & \multicolumn{2}{l}{{After LH}} \\
 \hline
 &  Precursor &  Precursor          &     Hydride     \\
 \hline
noise & $1.22\cdot10^4$ &    $4.16\cdot10^5$       &   $4.16\cdot10^5$       \\
signal & $7.27\cdot10^{9}$ &      $9.41\cdot10^{10}$     &     $2.15\cdot10^{12}$     \\
$N_{scans}$ & 2048 &    256000       &    256000      \\
$f_0~(MHz) $ & 45.02 &    45.072       &     45.072     \\
$\Delta f ~(MHz)$ & 1 &     0.33      &     0.33     \\
$\left( \frac{SNR}{\sqrt{N_{scans}}} \right)_{rn}$ & 13167 &    774       & 17802        
\end{tabular}
\endgroup

\subsection*{$CuH_{0.15}$}
\centering
\begingroup
\setlength{\tabcolsep}{12pt} 
\renewcommand{\arraystretch}{2.5} 
\begin{tabular}{l|l|l|l}
 & Ambient  & \multicolumn{2}{l}{{After LH}} \\
 \hline
 &  Precursor &  Precursor          &     Hydride     \\
 \hline
noise & $6.61\cdot10^4$ &    $4.37\cdot10^4$       &   $4.37\cdot10^4$       \\
signal & $4.12\cdot10^{10}$ &      $2.91\cdot10^{9}$     &     $5.05\cdot10^{9}$     \\
$N_{scans}$ & 2048 &    25600       &    25600      \\
$f_0~(MHz) $ & 45.061 &    45.017       &     45.017     \\
$\Delta f ~(MHz)$ & 2 &     0.33      &     0.33     \\
$\left( \frac{SNR}{\sqrt{N_{scans}}} \right)_{rn}$ & 13789 &    1020       & 1768        
\end{tabular}
\endgroup

\subsection*{$H_6S_5$}
\centering
\begingroup
\setlength{\tabcolsep}{12pt} 
\renewcommand{\arraystretch}{2.5} 
\begin{tabular}{l|l|l|l}
 & Ambient  & \multicolumn{2}{l}{{After LH}} \\
 \hline
 &  Precursor &  Precursor          &     Hydride     \\
 \hline
noise & $2.22\cdot10^9$ &    $4.46\cdot10^4$       &   $4.46\cdot10^4$       \\
signal & $1.26\cdot10^{15}$ &      $3.31\cdot10^{11}$     &     $1.78\cdot10^{10}$     \\
$N_{scans}$ & 103 &    3652       &    3652      \\
$f_0~(MHz) $ & 300.147 &    300.147       &     300.147     \\
$\Delta f ~(MHz)$ & 0.5 &     3.33      &     3.33     \\
$\left( \frac{SNR}{\sqrt{N_{scans}}} \right)_{rn}$ & 54685 &    47619       & 8532        
\end{tabular}
\endgroup

\subsection*{$YH_3$}
\centering
\begingroup
\setlength{\tabcolsep}{12pt} 
\renewcommand{\arraystretch}{2.5} 
\begin{tabular}{l|l|l|l}
 & Ambient  & \multicolumn{2}{l}{{After LH}} \\
 \hline
 &  Precursor &  Precursor          &     Hydride     \\
 \hline
noise & $2.357\cdot10^4$ &    $3.13\cdot10^4$       &   $3.13\cdot10^4$       \\
signal & $7.72\cdot10^{9}$ &      $7.38\cdot10^{9}$     &     $4.28\cdot10^{9}$     \\
$N_{scans}$ & 1024 &    1024       &    1024      \\
$f_0~(MHz) $ & 300.147 &    300.147       &     300.147     \\
$\Delta f ~(MHz)$ & 1 &     1      &     1     \\
$\left( \frac{SNR}{\sqrt{N_{scans}}} \right)_{rn}$ & 10240 &    7386       & 4279       
\end{tabular}
\endgroup

\subsection*{$YH_2$}
\centering
\begingroup
\setlength{\tabcolsep}{12pt} 
\renewcommand{\arraystretch}{2.5} 
\begin{tabular}{l|l|l|l}
 & Ambient  & \multicolumn{2}{l}{{After LH}} \\
 \hline
 &  Precursor &  Precursor          &     Hydride     \\
 \hline
noise & $2.357\cdot10^4$ &    $3.13\cdot10^4$       &   $3.13\cdot10^4$       \\
signal & $7.72\cdot10^{9}$ &      $7.38\cdot10^{9}$     &     $3.471\cdot10^{9}$     \\
$N_{scans}$ & 1024 &    1024       &    1024      \\
$f_0~(MHz) $ & 300.147 &    300.147       &     300.147     \\
$\Delta f ~(MHz)$ & 1 &     1      &     1     \\
$\left( \frac{SNR}{\sqrt{N_{scans}}} \right)_{rn}$ & 10240 &    7386       & 3479       
\end{tabular}
\endgroup

\end{document}